\documentclass[revtex4]{emulateapj}
\pdfoutput=1

\usepackage{amsmath}
\usepackage{color}
\usepackage{xspace}
\usepackage{lscape}
\usepackage{hyperref}
\usepackage{textcomp}

\newcommand{\ee}[1]{\!\times\!10^{#1}}
\newcommand{\gws}{gravitational waves\xspace}
\newcommand{\gw}{gravitational wave\xspace}

\newcommand{\dms}[4]{$#1$\degr#2\arcmin#3\farcs#4}
\newcommand{\hms}[4]{#1$^{\mathrm h}$#2$^{\mathrm m}$#3\fs#4}
\newcommand{\beq}{\begin{equation}}
\newcommand{\eeq}{\end{equation}}
\newcommand{\F}{\ensuremath{\mathcal{F}}}
\newcommand{\G}{\ensuremath{\mathcal{G}}}
\newcommand{\FG}{\F/\G}
\newcommand{\fgw}{\ensuremath{f_{\rm gw}}\xspace}
\newcommand{\frot}{\ensuremath{f_{\rm rot}}\xspace}

\begin{document}

\keywords{gravitational waves - pulsars: general}

\title{Gravitational waves from known pulsars: results from the initial detector era}
\shorttitle{Gravitational waves from known pulsars}

% $Id: authors_list_LVC.tex,v 1.9 2013/09/30 13:42:04 mpitkin Exp $
% authorlist from LSC-VIRGO Joint Author List, LSC February 2013 - Virgo July 2013 https://dcc.ligo.org/LIGO-M1300495

\author{%
J.~Aasi\altaffilmark{1},
J.~Abadie\altaffilmark{1},
B.~P.~Abbott\altaffilmark{1},
R.~Abbott\altaffilmark{1},
T.~Abbott\altaffilmark{2},
M.~R.~Abernathy\altaffilmark{1},
T.~Accadia\altaffilmark{3},
F.~Acernese\altaffilmark{4,5},
C.~Adams\altaffilmark{6},
T.~Adams\altaffilmark{7}
% ...
}

\altaffiltext{1}{LIGO - California Institute of Technology, Pasadena, CA 91125, USA}
\altaffiltext{2}{Louisiana State University, Baton Rouge, LA 70803, USA}
\altaffiltext{3}{Laboratoire d'Annecy-le-Vieux de Physique des Particules (LAPP), Universit\'e de Savoie,
CNRS/IN2P3, F-74941 Annecy-le-Vieux, France}
\altaffiltext{4}{INFN, Sezione di Napoli, Complesso Universitario di Monte S.Angelo, I-80126 Napoli, Italy}
\altaffiltext{5}{Universit\`a di Salerno, Fisciano, I-84084 Salerno, Italy}
\altaffiltext{6}{LIGO - Livingston Observatory, Livingston, LA 70754, USA}
\altaffiltext{7}{Cardiff University, Cardiff, CF24 3AA, United Kingdom}
% ...
\affil{The LIGO Scientific Collaboration \& The Virgo Collaboration}

% $Id: authors_pulsar.tex,v 1.13 2013/09/18 09:56:11 mpitkin Exp $
% author list of pulsar astronomers

\author{
S.~Buchner\altaffilmark{127,128},
I.~Cognard\altaffilmark{129,130},
A.~Corongiu\altaffilmark{131},
N.~D'Amico\altaffilmark{131,132},
C.~M.~Espinoza\altaffilmark{133,134},
P.~C.~C.~Freire\altaffilmark{135},
E.~V.~Gotthelf\altaffilmark{28},
L.~Guillemot\altaffilmark{135},
J.~W.~T.~Hessels\altaffilmark{136,137},
G.~B.~Hobbs\altaffilmark{138},
M.~Kramer\altaffilmark{133,135},
A.~G.~Lyne\altaffilmark{133},
F.~E.~Marshall\altaffilmark{37},
A.~Possenti\altaffilmark{131},
S.~M.~Ransom\altaffilmark{139},
P.~S.~Ray\altaffilmark{140},
J.~Roy\altaffilmark{141},
and B.~W.~Stappers\altaffilmark{133}
}

\altaffiltext{127}{Hartebeesthoek Radio Astronomy Observatory, PO Box 443, Krugersdorp, 1740,  South Africa}
\altaffiltext{128}{School of Physics, University of the Witwatersrand, Johannesburg, South Africa}
\altaffiltext{129}{LPC2E/CNRS-Universit\'e d'Orl\'eans, 45071 Orl\'eans, France}
\altaffiltext{130}{Nan\c{c}ay/Paris Observatory, 18330 Nan\c{c}ay, France}
\altaffiltext{131}{INAF - Osservatorio Astronomico di Cagliari, Poggio dei Pini, 09012 Capoterra, Italy}
\altaffiltext{132}{Dipartimento di Fisica Universit\`a di Cagliari, Cittadella Universitaria, I-09042 Monserrato, Italy}
\altaffiltext{133}{Jodrell Bank Centre for Astrophysics, School of Physics and Astronomy, University of Manchester,
Manchester M13 9PL, UK}
\altaffiltext{134}{Instituto de Astrof\'isica, Facultad de F\'isica, Pontificia Universidad 
Cat\'olica de Chile, Casilla 306, Santiago 22, Chile}
\altaffiltext{135}{Max-Planck-Institut f\"{u}r Radioastronomie, Auf dem H\"{u}gel 69, D-53121 Bonn, Germany}
\altaffiltext{136}{ASTRON, the Netherlands Institute for Radio Astronomy, Postbus 2, 7990 AA, Dwingeloo, The 
Netherlands}
\altaffiltext{137}{Astronomical Institute ``Anton Pannekoek'', University of Amsterdam, Science Park 904, 1098 XH
Amsterdam, The Netherlands}
\altaffiltext{138}{Australia Telescope National Facility, CSIRO, PO Box 76, Epping NSW 1710, Australia}
\altaffiltext{139}{National Radio Astronomy Observatory, Charlottesville, VA 22903, USA}
\altaffiltext{140}{Space Science Division, Naval Research Laboratory, Washington, DC 20375-5352, USA}
\altaffiltext{141}{National Centre for Radio Astrophysics, Pune 411007, India}

% from LVC list
\altaffiltext{\dag}{Deceased, April 2012.}%  
\altaffiltext{\ddag}{Deceased, May 2012.}%

\begin{abstract}
We present the results of searches for gravitational waves from a large selection of pulsars using data from the most
recent science runs (S6, VSR2 and VSR4) of the initial generation of interferometric gravitational wave detectors LIGO
(Laser Interferometric Gravitational-wave Observatory) and Virgo. We do not see evidence for gravitational wave emission
from any of the targeted sources but produce upper limits on the emission amplitude. We highlight the results from seven
young pulsars with large spin-down luminosities. We reach within a factor of five of the canonical spin-down limit for
all seven of these, whilst for the Crab and Vela pulsars we further surpass their spin-down limits. We present new or
updated limits for 172 other pulsars (including both young and millisecond pulsars). Now that the detectors are
undergoing major upgrades, and, for completeness, we bring together all of the most up-to-date results from all pulsars
searched for during the operations of the first-generation LIGO, Virgo and GEO600 detectors. This gives a total of 195
pulsars including the most recent results described in this paper.
\end{abstract}

\maketitle

%%%%%%%%%%%%%%%%%

\section{Introduction}\label{sec:intro}
Pulsars are spinning, magnetized neutron stars with slowly decreasing rotation rates. In the model of a triaxial
ellipsoid star, a deformation (possibly from shear strains in the solid part(s) of the star, or from magnetic stresses)
can appear as a time-varying quadrupole moment as the star rotates. The observed loss of rotational energy, known as
the spin-down luminosity (given by $\dot{E} = I_{zz}\Omega|\dot{\Omega}| = 4\pi^2I_{zz}\frot|\dot{f}_{\rm rot}|$, where
$I_{zz}$ is the moment of inertia around the principal axis (aligned with the rotation axis), \frot is the rotation
frequency, and $\dot{f}_{\rm rot}$ is the rotational frequency derivative) provides a huge reservoir of energy. Along
with magnetic dipole radiation some fraction of this reservoir is potentially dissipated through \gw emission
(see \citealp{Shklovskii:1969, Ostriker:1969, Ferrari:1969, Melosh:1969} for four contemporaneous calculations of \gw
emission from soon after pulsars were discovered, or e.g.\ \citealp{Owen:2006} for a review of more recent emission
mechanism calculations). Known pulsars usually have precisely determined frequency evolutions and sky-positions making
them ideal targets for \gw detectors. If a pulsar is monitored regularly through electromagnetic observations it can
yield a coherent phase model, which allows \gw data to be coherently integrated over months or years.

Since the initial science data runs of the Laser Interferometric Gravitational-wave Observatory (LIGO), Virgo and
GEO600, searches have been performed for continuous quasi-monochromatic \gw emission from many known pulsars
\citep{Abbott:2004, Abbott:2005, Abbott:2007a, Abbott:2008, Abbott:2010, Abadie:2011}. Most recently 116 known pulsars
were targeted using data from LIGO's fifth science run \citep[S5][]{Abbott:2010}, and the Vela pulsar
(J0835\textminus4510) was targeted using data from Virgo's second science run (VSR2). These searches reported no
detections, but provided upper limits on the \gw amplitude from the sources and surpassed the so-called spin-down limit
(see Section~\ref{sec:signal}) for the Crab and Vela pulsars.

We aim here to search for \gw emission from a large selection of stars including some of those with the largest
spin-down luminosities. Due to the sensitivity reduction caused at low frequency by seismic noise at the detectors, 
it is not worthwhile to search for pulsars with rotational frequencies, \frot, smaller than about 10\,Hz, which
corresponds to \gw mass quadrupole emission at frequencies, \fgw = 2\frot, smaller than 20\,Hz. The exact value of this
\gw low-frequency cut-off is rather arbitrary, our choice of taking 20\,Hz is motivated by the presence of several noise
lines and bumps in Virgo data at lower frequencies. In general, young pulsars, with large spin-down luminosities are
searched for at lower frequencies where the Virgo detector has better sensitivity, whereas the search for millisecond
pulsars (MSPs) is conducted at higher frequencies where the LIGO detectors are more sensitive. The selection of pulsars
will be discussed more fully in Section~\ref{sec:pulsars}.

\subsection{The signal}\label{sec:signal}
The expected quadrupolar \gw signal from a triaxial neutron star\footnote{We use `triaxial neutron star' as shorthand
for a star with some asymmetry with respect to its rotation axis and therefore a triaxial moment of inertia ellipsoid.}
steadily spinning about one of its principal axes of inertia is at twice the rotation frequency, with a strain of
\begin{align}
h(t) = &\frac{1}{2}F_+(t, \psi)h_0(1+\cos{}^2\iota)\cos{\phi(t)} \nonumber \\
& + F_{\times}(t, \psi)h_0\cos{\iota}\sin{\phi(t)}
\end{align}
in the detector, where
\begin{equation}\label{eq:h0}
h_0=\frac{16\pi^2G}{c^4}\frac{I_{zz}\varepsilon \frot^2}{d}
\end{equation}
is the dimensionless \gw strain amplitude. $h_0$ is dependent on $I_{zz}$, the fiducial equatorial ellipticity, defined
as $\varepsilon=\frac{I_{xx}-I_{yy}}{I_{zz}}$ in terms of principal moments of inertia, the rotational frequency,
$\frot$, and the distance to the source $d$. The signal amplitudes in the two polarizations (`+' and `$\times$') depend
on the inclination of the star's rotation axis to the line-of-sight, $\iota$, while the detector antenna pattern
responses for the two polarization states, $F_+(t, \psi)$ and $F_\times(t, \psi)$, depend on the \gw polarization angle,
$\psi$, as well as the detector location, orientation and source sky position. The \gw phase evolution, $\phi(t)$,
depends on both the intrinsic rotational frequency and frequency derivatives of the pulsar and on Doppler and
propagation effects. These extrinsic effects include relativistic modulations caused by the Earth's orbital and
rotational motion, the presence of massive bodies in the solar system close to the line-of-sight to the pulsar, the
proper motion of the pulsar, and (in the case of a binary system) pulsar orbital motions. We will assume that $\phi(t)$
is phase-locked to the electromagnetic pulse phase evolution, but with double the value and with an initial phase
offset, $\phi_0$. Given this phase evolution, the four unknown search parameters are simply $h_0$, $\cos{\iota}$,
$\phi_0$ and $\psi$. The \gw amplitude is related to the star's $l=m=2$ mass quadrupole moment via \citep[see
e.g.][]{Owen:2005}
\begin{equation}\label{eq:q22}
Q_{22} = \sqrt{\frac{15}{8\pi}} I_{zz}\varepsilon = h_0\left( \frac{c^4 d}{16\pi^2G \frot^2} \right)
\sqrt{\frac{15}{8\pi}},
\end{equation}
where $Q_{22}$ is the slightly non-standard definition of quadrupole moment used in \citet{Ushomirsky:2000} and many
subsequent papers. This value can be constrained independently of any assumptions about the star's equation of state and
moment of inertia.

If we allocate all the spin-down luminosity, $\dot{E}$, to \gw luminosity, $\dot{E}_{\rm gw}$, where
\begin{align}\label{eq:esd}
\dot{E}_{\rm gw} =& \frac{2048\pi^6}{5}\frac{G}{c^5}\frot^6(I_{zz}\varepsilon)^2, \nonumber \\
 =& \frac{8\pi^2}{5}\frac{c^3}{G}\frot^2 h_0^2 d^2,
\end{align}
then we have the canonical `spin-down limit' on \gw strain\footnote{As noted in \citet{Johnson-McDaniel:2013}, the
versions of this equation given inline in the first paragraph of \citet{Abbott:2008}, as equation~(1) in
\citet{Abbott:2010} and as equation~(14) in \citet{Abadie:2011} are incorrect and should have $I_{38}$ substituted for
$I_{38}^{1/2}$.}
\begin{align}\label{eq:h0sd}
h_0^{\rm sd} = &\left(\frac{5}{2} \frac{G I_{zz} \dot{f}_{\rm rot}}{c^3 d^2 \frot}
\right)^{1/2}\nonumber \\
 = & \> 8.06\times 10^{-19}\frac{I^{1/2}_{38}}{d_{\rm kpc}}\left(\frac{|\dot{f}_{\rm rot}|}{\frot}\right)^{1/2},
\end{align}
where $I_{38}$ is the star's moment of inertia in the units of $10^{38}$\,kg\,m$^2$, and $d_{\rm kpc}$ is the distance
to the pulsar in kiloparsecs. The spin-down limit on the signal amplitude corresponds (via equation~\ref{eq:h0}) to an
upper limit on the star's fiducial ellipticity\footnote{Again, as noted in \citet{Johnson-McDaniel:2013}, the versions
of this equation given inline in Section~3 of \citet{Abbott:2008} and as equation~(7) in \cite{Abbott:2010} are
incorrect and should have $I_{38}$ substituted for $I_{38}^{-1}$.}
\begin{equation}\label{eq:eps_sd}
\varepsilon^{\rm sd} = 0.237\left(\frac{h_0^{\rm sd}}{10^{-24}}\right)\frot^{-2} I_{38}^{-1} d_{\rm kpc}.
\end{equation}
\citet{Johnson-McDaniel:2013} shows how to relate this to the physical ellipticity of the star's surface for a given
equation of state.

A \gw strain upper limit that is below the spin-down limit is an important milestone, as such a measurement is probing
uncharted regions of the parameter space. Likewise it directly constrains the fraction of spin-down power that could be
due to the emission of \gws, which gives insight into the overall spin-down energy budget.

\subsection{The science runs}\label{sec:runs}
In this paper we have used calibrated data from the Virgo second \citep{Aasi:2012} and fourth science runs (VSR2 and
VSR4) and the LIGO sixth science run (S6). Virgo's third science run (VSR3) was relatively insensitive in
comparison with VSR4 and has not been included in this analysis. This was partially because Virgo introduced monolithic
mirror suspensions prior to VSR4 which improved sensitivity in the low-frequency range. During S6, the two LIGO 4\,km
detectors at Hanford, Washington (LHO/H1), and Livingston, Louisiana (LLO/L1), were running in an enhanced configuration
\citep{Adhikari:2006} over that from the previous S5 run \citep{Abbott:2009b}. Table~\ref{tab:runs} shows dates of
the runs, the duty factors and science data lengths for each detector that we analyzed.

\begin{deluxetable*}{cccc}
\tablewidth{0pt}
\tabletypesize{\footnotesize}
\tablecaption{Science runs. \label{tab:runs}}
\tablehead{\colhead{Run} & \colhead{Dates} & \colhead{Duty factor ($\%$)} & \colhead{Data length (days)}}
\startdata
VSR2 & 2009 Jul 7 (20:55 UTC) -- 2010 Jan 8 (22:00 UTC) & 80.4 & 149 \\\
VSR4 & 2011 Jun 3 (10:27 UTC) -- 2011 Sep 5 (13:26 UTC) & 81.0 & 76 \\
S6 Hanford (H1) & 2009 Jul 7 (21:00 UTC) -- 2010 Oct 21 (00:00 UTC) & 50.6 & 238 \\
S6 Livingston (L1) & 2009 Jul 7 (21:00 UTC) -- 2010 Oct 21 (00:00 UTC) & 47.9 & 225
\enddata
\end{deluxetable*}

The Virgo and LIGO data used in these analyses have been calibrated through different reconstruction procedures, but
both based ultimately on the measured response to actuation controls on the positions of the mirrors that define the
interferometers. For Virgo VSR2, the calibration uncertainty was about 5.5\% in amplitude and $\sim$50\,mrad (3\degr) in
phase over most of the frequency range \citep{Accadia:2011}. For VSR4, the uncertainty amounted to about 7.5\% in
amplitude and to ($40+50f$)\,mrad in phase, where $f$ is the frequency in kilohertz, for frequencies up to 500\,Hz
\citep{Mours:2011}. For LIGO, the S6 calibration uncertainties over the  relevant frequency range (50--1500\,Hz) were up
to $\sim$19\% in amplitude and $\sim$170\,mrad (10\degr) in phase for L1, and up to $\sim$16\% in amplitude and
$\sim$120\,mrad (7\degr) for H1 \citep{Bartos:2011}. These phase errors are well within the range \citep[i.e.\ less than
30\degr as applied in][]{Abbott:2007a} that would cause significant loss in signal power due to decoherence between the
pulsar signal and the assumed phase evolution.

\subsection{Methods}\label{sec:methods}
We used three semi-independent methods \citep[very similar to those used in the Vela pulsar search in][]{Abadie:2011}
to search for signals described in Section~\ref{sec:signal}. Here, we briefly outline their operation, but for full
descriptions we refer the reader to the references below. Two of the search methods work with time domain data that has
been heterodyned to remove the signal's phase evolution and then heavily decimated. This leaves a complex data stream in
which any signal would only be modulated by the detector's beam pattern. In the first method, this data stream is used
to give Bayesian parameter estimates of the unknown signal parameters\footnote{For this analysis the parameter posterior
distributions were recreated using a Markov chain Monte Carlo \citep{Abbott:2010}. For each pulsar five independent
chains were produced with 50\,000 burn-in samples and 200\,000 posterior samples in each. The chains were thinned using
the autocorrelation length to give uncorrelated samples, and to test for convergence, the chains were then examined by
eye, and a Gelman-Rubins test was performed \citep[see e.g.][]{Brooks:1998}.} \citep{Dupuis:2005}. The second method
computes the maximum likelihood \F-statistic rather than a Bayesian posterior (or in case where $\psi$ and $\iota$ are
well constrained, the \G-statistic) \citep{Jaranowski:2010}. The third method \citep{Astone:2010} makes use of a Short
Fourier Transform Database (SFDB) of each detector's data. After the extraction of a small frequency band around the
signal's expected frequency, the Doppler effect, Einstein delay and spin-down are removed in the time domain and the
data are down-sampled with a re-sampling technique. Two matched filters on the `+' and `$\times$' signal Fourier
components are then computed at the five frequencies at which the signal power is spread due to the signal amplitude and
phase modulation; they are used to build a detection statistic and to estimate signal parameters in the case of
detection. This 5-vector method has been extended over that used in \citet{Abadie:2011} to allow for coherent analysis
of data from multiple detectors \citep{Astone:2012}. Each of these methods can incorporate prior information on the
pulsar's inclination and polarization angle. From here on, we will refer to the first method as the {\it Bayesian}
method\footnote{For this analysis, the results were produced with version 6.16 of the LSC Algorithm Library Suite
(LALSuite) \url{https://www.lsc-group.phys.uwm.edu/daswg/projects/lalsuite.html}.}, the second as the {\it
\FG-statistic} method and the third as the {\it $5n$-vector} method, where $n$ refers to the number of datasets
coherently combined.

All three methods apply some data cleaning. The procedure used to obtain the heterodyned data removes extreme outliers
by running two passes of a scheme that identifies points with absolute values greater than five times the standard
deviation of the dataset. The \FG-statistic method performs further cleaning of this data through the Grubbs test
\citep[see][]{Abadie:2011}. In the $5n$-vector method, after an initial time-domain cleaning before the construction of
the SFDB, a further cleaning step is applied on the final down-sampled time series in which the largest outliers
belonging to the non-Gaussian tail of the data amplitude distribution are removed.

We have incorporated some limits from the previous LIGO S5 results \citep{Abbott:2010} as priors in the Bayesian
analysis. However, the S6/VSR2,4 phase models were produced with updated pulsar ephemerides resulting in an unknown
phase offset between them and the S5 results. We have, therefore, simply used the S5 marginalised posterior on $h_0$ and
$\cos{\iota}$, $p(h_0,\cos{\iota})$, as our prior for the new results.  In the case of glitching pulsars (see
Section~\ref{sec:pulsars}), we used the same approach and (incoherently) combined the separate coherent analyses
produced between glitches. In the case of the \FG-statistic method, the results from different detectors or
different inter-glitch periods are combined incoherently by adding the respective statistics. Also, for the $5n$-vector
method, results from different inter-glitch periods are incoherently combined by summing the corresponding statistics.
Our reasons for not coherently combining the data over glitches are twofold. The first is that we do not know how a
glitch would effect the relative phase offset between the electromagnetic pulses and the \gw signal. The second reason
is a practical consideration based on the timing solutions we have for our pulsars. For three of the four glitching
pulsars (all except J0537\textminus6910) in this analysis we have separate timing solutions for each inter-glitch
period. These separate timing solutions, as provided by the pulsar timing software TEMPO(2), do not give an epoch
defined at a fixed pulse phase (i.e.\ the epoch is not given as the time of the peak of a particular pulse), so there is
some unknown phase offset between the separate solutions. However, if this phase offset were known (e.g.\ by going back
to the original pulsar pulse time of arrival data) the gain in sensitivity would still be minimal: for the Vela pulsar
and J1813\textminus1246 the data from VSR4 (which was after the glitches in these pulsars) was much more sensitive, so
completely dominates the result; for J1952+3252 the post-glitch data contains the latter part of S6, which was more
sensitive and would again dominate the results. For J0537\textminus6910 the epoch for each inter-glitch timing solution
is defined (see Table~\ref{tab:J0537-6910ephem}), so the electromagnetic phase could be tracked over the glitches, but
again our result is dominated by the longest and most sensitive inter-glitch period.

Even without a detection, all three methods can be used to produce upper limits on the \gw amplitude from the pulsars.
Here, we will quote 95\% confidence upper limits on the amplitude. In the Bayesian method, an upper limit on the $h_0$
posterior (after marginalization over the orientation parameters) is found by calculating the upper bound, from zero, on
the integral over this posterior that encloses 95\% of the probability. In the \FG-statistic method, a frequentist upper
limit is calculated through Monte-Carlo simulations, which find the value of $h_0$ for which 95\% of trials exceed the
maximum likelihood statistic\footnote{The \FG-statistic is most suitable for signal detection, whilst the upper limit
derived from it here is mainly given for completeness. In the future a more sophisticated method, such as that of
\citet{Feldman:1998}, may be used to produce frequentist confidence intervals for this analysis.}. The $5n$-vector
method computes an upper limit on the $H_0$ posterior, given the actual value of the detection statistic, and the
marginalization over the other parameters is implicitly done in the Monte Carlo simulation used to compute the
likelihood. The amplitude, $H_0$, is linked to the classical $h_0$, given by equation~\ref{eq:h0}, by the relation
$H_0=\frac{h_0}{2}\sqrt{1+6\cos{}^2\iota+\cos{}^4\iota}$ \citep[see equation~(A5) in][]{Abadie:2011}. $H_0$ is the
strain amplitude of a linearly polarized signal with polarization angle $\psi=0$. In order to convert an upper limit on
$H_0$ to an upper limit on $h_0$, we use the previous equation replacing the coefficient on the right hand side with its
mean value over the distribution of $\cos{\iota}$ used in the upper limit procedure. This is justified by the fact that
the posterior distribution of $H_0$ is not dependent on $\iota$. The three methods have been tested with hardware and
software simulated signal injections to check that they can recover the expected signal model \citep[see
e.g.][]{Abadie:2011}. In the Bayesian analysis these upper limits are really 95\% credibility, or degrees-of-belief,
values, whereas for the frequentist analysis these are 95\% confidence values. These are both asking different questions
and in general should not be expected to produce identical results. A brief discussion of this is given in the first
search for a pulsar in LIGO data in \citet{Abbott:2004}, whilst a more technical discussion of the differences between
the upper limits can be found in \citet{Roever:2011}.

\section{Pulsar selection}\label{sec:pulsars}
The sensitivity of the Virgo and LIGO detectors allows us to target pulsars with $\frot > 10$\,Hz. Currently the
Australia Telescope National Facility (ATNF) pulsar catalog \citep{Manchester:2005} contains data for 368 pulsars (out
of a total of 2264) consistent with this criterion\footnote{ATNF pulsar catalog v1.47
\url{http://www.atnf.csiro.au/people/pulsar/psrcat/}}. The majority of these ($\sim$90\%) are recycled MSPs that have
been spun-up to high rotation frequencies by accretion from a binary companion which may still be present \citep[see
e.g.][for an overview of MSPs and binary pulsars]{Lorimer:2008}. MSPs spin down slowly (with $\dot{f}_{\rm rot}$ between
approximately $-10^{-14}$ and $-10^{-17}$\,Hz/s) and have characteristic ages\footnote{Characteristic age is given by
$\tau = -(1/(n-1))(\frot/\dot{f}_{\rm rot})$, which, for a magnetic dipole braking index of $n=3$, gives $\tau = -f_{\rm
rot}/(2\dot{f}_{\rm rot})$, and for purely \gw (quadrupole) spin-down would be $n=5$, giving $\tau =
-\frot/(4\dot{f}_{\rm rot})$ \citep[a ``gravitar'',][]{Palomba:2005,Knispel:2008}.} greater than a few times $10^8$
years, implying a comparatively weak surface polar magnetic field ($10^8 \lesssim B_s \lesssim 10^9$\,G, via the
relation for an orthogonal rotator with radius 10\,km and $I_{zz} = I_{38}$ of $B_s =
3.3\ee{19}(|\dot{f}_{\rm rot}|/\frot^3)^{1/2}$\,G) compared to ``normal'' pulsars. About 10\% are young pulsars with
$\dot{f}_{\rm rot}$ between approximately $-10^{-10}$ and $-10^{-12}$\,Hz/s, characteristic ages of between $\sim$1000
and a few tens of thousands of years, and therefore with the large implied surface magnetic fields of ``normal''
pulsars, $B_s \sim10^{12}$\,G. They are situated towards the low-frequency end of our sensitivity range.

Young pulsars have large spin-downs and relatively low frequencies, so in general have the highest \gw spin-down limits,
see equation~\ref{eq:h0sd}. This makes them particularly important targets as the limits can be within reach of current
detectors. Equations~\ref{eq:q22} and \ref{eq:eps_sd} show that to produce emission at around the spin-down limit the
required mass quadrupole/ellipticity would have to be large, at a level consistent with only the most exotic neutron
star equations of state (see the discussion in Section~\ref{sec:discussion}). Such strong emission is unlikely, but its
detection would have profound implications. Young pulsars also often show rotational anomalies such as glitches and
timing noise \citep[see e.g.][]{Helfand:1980}. The underlying causes of such phenomena are still quite uncertain, and
\gw data would be a powerful constraint. For the MSPs, the spin-down limits are generally several orders of magnitude
below those for the young pulsars. They are located, however, in a more sensitive frequency range.

\subsection{Electromagnetic pulsar observations}\label{sec:em}
For this analysis, we have obtained ephemerides using radio, X-ray and $\gamma$-ray observations. The radio telescope
observations have come from a variety of sources: the 12.5-m telescope and Lovell telescope at Jodrell Bank in the UK,
the 26-m telescope at Hartebeesthoek in South Africa, the 15-m eXperimental Development Model (XDM) telescope in South
Africa, the Giant Metrewave Radio Telescope (GMRT) in India, the Robert C.\ Byrd Green Bank Radio Telescope (GBT) in the
US, the Parkes radio telescope in Australia, the Nan\c{c}ay Decimetric Radio Telescope in France and the Hobart radio
telescope in Australia. High energy X-ray and $\gamma$-ray timings have been obtained from the Rossi X-ray Timing
Explorer (RXTE) and the {\it Fermi} Large Area Telescope (LAT).

In total, for this analysis, we collected timing solutions for 179 pulsars. This selection includes 73 pulsars that have
not been previously studied. However, for five of the pulsars targeted in the S3/S4 analysis \citep{Abbott:2007a} and
another eleven of the pulsars targeted in the S5 analysis \citep{Abbott:2010}, new coherent timing solutions were not
available, so these stars\footnote{The five additional pulsars targeted in S3/S4 were J1435\textminus6100,
J1629\textminus6902, J1757\textminus5322, J1911+0101A and J1911+0101B and the eleven additional pulsars targeted in S5
were J1701\textminus3006B, J1701\textminus3006C, J1748\textminus2446P, J1748\textminus2446ad, J1824\textminus2452B,
J1824\textminus2452C, J1824\textminus2452E, J1824\textminus2452F, J1824\textminus2452G, J1824\textminus2452H,
J1824\textminus2452J.} have not been included in this search.

\subsubsection{High interest targets}
As discussed in \citet{Abbott:2008}\footnote{Note that \citet{Johnson-McDaniel:2013} computes even larger potential
moments of inertia at $\sim$$5\ee{38}$\,kg\,m$^2$ for some solid quark stars.}, due to our ignorance of the correct
neutron star equation of state there is a large uncertainty in the moments of inertia for our targets, from $1$ to
$3\ee{38}$\,kg\,m$^2$. Therefore, the canonical spin-down limit estimates could be increased by a factor of $\sim$1.7.
Also, there are uncertainties in some pulsar distance measurements of up to a factor of two which could further increase
or decrease the spin-down limit. We therefore identified all sources that were within a factor of four of the canonical
spin-down limit as worthy of special attention. Seven of the pulsars for which we have obtained timing solutions beat,
or approach to within a factor of four, this limit. The electromagnetic observation epochs for each pulsar (which
include each inter-glitch epoch for pulsars that glitched during the analysis) are given in Table~\ref{tab:pulparams}.
\begin{deluxetable}{c}
\tablewidth{0pt} \tabletypesize{\footnotesize} \tablecaption{Electromagnetic observation epochs for the high interest
pulsars. \label{tab:pulparams}} \tablehead{\colhead{MJD and date}}
\startdata
\cutinhead{J0534+2200 (Crab pulsar)}
54997 (2009 Jun 15) -- 55814 (2011 Sep 10) \\
\cutinhead{J0537\textminus6910 (N157B)}
54897 (2009 Mar 7) -- 55041 (2009 Jul 29) \\
55045 (2009 Aug 2) -- 55182 (2009 Dec 17) \\
55185 (2009 Dec 20) -- 55263 (2010 Mar 8) \\
55275 (2010 Mar 20) -- 55445 (2010 Sep 6) \\
55458 (2010 Sep 19) -- 55503 (2010 Nov 3) \\
\cutinhead{J0835\textminus4510 (Vela pulsar)}
54983 (2009 Jun 1) -- 55286 (2010 Mar 31) \\
55713 (2011 Jun 1) -- 55827 (2011 Sep 23) \\
\cutinhead{J1813\textminus1246}
54693 (2008 Aug 15) -- 55094 (2009 Sep 20) \\
55094 (2009 Sep 20) -- 55828 (2011 Sep 24) \\
\cutinhead{J1833\textminus1034 (G21.5\textminus0.9)}
55041 (2009 Jul 29) -- 55572 (2011 Jan 11) \\
\cutinhead{J1913+1011}
54867 (2009 Feb 5) -- 55899 (2011 Dec 4) \\
\cutinhead{J1952+3252 (CTB 80)}
54589 (2008 May 3) -- 55325 (2010 May 9) \\
55331 (2010 May 15) -- 55802 (2011 Aug 29) \\
\enddata
\end{deluxetable}
Further details of these observations are given below:
\begin{description}
\item[J0534+2200 (the Crab pulsar)] We have used the Jodrell Bank Monthly Ephemeris \citep{Lyne:1993} to track the phase
of the Crab pulsar over the period of our runs. This ephemeris has timing solutions using the DE200 solar system
ephemeris and the TDB time coordinate system. During S6/VSR2,4 the pulsar did not show signs of any timing glitches.
\item[J0537\textminus6910 (N157B)] Long-term X-ray timing has been performed with the RXTE \citep{Middleditch:2006}.
Recent data covering S6 shows four glitches over the span of our science runs and the ephemerides for each inter-glitch
epoch are given in Appendix~\ref{app:j0537}. The timing solutions used the DE200 solar system ephemeris
\citep[see][]{Marshall:1998} and the TDB time coordinate system. Several more glitches have been observed since the end
of our science runs, but we do not report on them here.
\item[J0835\textminus4510 (the Vela pulsar)] Radio observations over the period of VSR2 were taken with the Hobart radio
telescope in Tasmania and the Hartebeesthoek 26-m radio telescope in South Africa \citep{Abadie:2011}. Radio timing over
the VSR4 run was performed with the XDM telescope and the 26-m telescope at Hartebeesthoek. The timing solutions have
used the DE405 solar system ephemeris and the TCB time coordinate system. Vela was observed to glitch on 2010 July 31
\citep{Buchner:2010}, between VSR2 and VSR4, but it has not glitched since then.
\item[J1813\textminus1246] This pulsar was discovered in a search of gamma-ray data from the {\it Fermi} LAT
\citep{Abdo:2009}, and using the unbinned maximum likelihood methods of \citet{Ray:2011} timing measurements were made
covering all our runs. It was observed to glitch once during this time on 2009 September 20. Pre-and-post glitch timing
solutions have been produced using the DE405 solar system ephemeris and the TDB time coordinate system.
\item[J1833\textminus1034 (G21.5\textminus0.9)] The period from the start of S6/VSR2 until 2011 January is covered by
observations made with the Giant Metrewave Radio Telescope (GMRT) \citep{Roy:2012}. During this period, one glitch was
observed, with a best fit epoch of 2009 November 6 (MJD 55142$\pm2$). To remove its effect, an ephemeris fit was
performed on timing data excluding 80 days after the glitch. The timing solution uses the DE405 solar system ephemeris
and the TDB time coordinate system.
\item[J1913+1011] This pulsar was observed at Jodrell Bank and showed no timing anomalies over the science runs. The
timing solution uses the DE405 solar system ephemeris and the TDB time coordinate system.
\item[J1952+3252 (CTB 80)] This pulsar was observed over the whole of our science runs at Nan\c{c}ay and Jodrell Bank.
It glitched on 2010 May 11 (MJD 55327), between the end of S6/VSR2 and the start of VSR4. Phase incoherent pre- and
post-glitch timing solutions have been produced using the DE405 solar system ephemeris and the TCB time ephemeris. The
solution include fits to the timing noise using the {\sc tempo2} {\sc Fitwaves} method described in \cite{Hobbs:2006}.
\end{description}

For several of these pulsars potential constraints on their orientations (the inclination $\iota$ and polarization angle
$\psi$\footnote{In \cite{Ng:2008} the inclination is denoted by $\zeta$ and the position angle $\Psi$ is equivalent to
our polarization angle. Our searches are insensitive to rotations of $90\degr$ in the polarization angle, so our quoted
values are rotated into the range $-45\degr < \psi < 45\degr$.}) are available from observations of their pulsar wind
nebulae \citep{Ng:2004, Ng:2008}. These are listed in Table~\ref{tab:constraints} where the uncertainties used are
estimated from the systematic and statistical values given in \cite{Ng:2004, Ng:2008}, and the mean angle value is used
if multiple fits are given (e.g.\ fits to the inner and outer tori of the Crab pulsar wind nebula). We briefly discussed
how these constraints are used in the analyses in Section~\ref{sec:methods}.

\begin{deluxetable}{lcc}
\tablecaption{Implied orientations of pulsars from their pulsar wind nebulae observations \citep{Ng:2004, Ng:2008}.
\label{tab:constraints}} \tablehead{\colhead{Pulsar} & \colhead{$\iota$} & \colhead{$\psi$}}
\startdata
J0534+2200 (Crab pulsar) & $62\degr\!.2 \pm 1\degr\!.9$ & $ 35\degr\!.2 \pm 1\degr\!.5$ \\
J0537\textminus6910 & $92\degr\!.8 \pm 0\degr\!.9$ & $41\degr\!.0 \pm 2\degr\!.2$ \\
J0835-4510 (Vela pulsar) & $63\degr\!.6 \pm 0\degr\!.6$ & $40\degr\!.6 \pm 0\degr\!.1$ \\
J1833\textminus1034 & $85\degr\!.4 \pm 0\degr\!.3$ & $45\degr \pm 1\degr$ \\
J1952+3252\tablenotemark{$\dagger$} & \nodata & $-11\degr\!.5 \pm 8\degr\!.6$
\enddata
\tablenotetext{$\dagger$}{The polarization angle is not taken from a fit to the pulsar wind nebula, but instead is the
average of the angle calculated from proper motion measurements and H$\alpha$ observations of a bow shock
\citep{Ng:2004}.}
\end{deluxetable}

For J0534+2200 and J0537\textminus6910, the Bayesian method also makes use of results from the LIGO S5 run
\citep{Abbott:2010} as a prior on the $h_0$ and $\cos{\iota}$ parameters. During S5, both of these pulsars glitched, and
the data for each inter-glitch period was analyzed independently. Results were also produced assuming that the data
could be analyzed coherently over the glitches. To avoid the assumptions about coherence over the glitches, we have used
the independent inter-glitch results that gave the lowest $h_0$ as the prior for the current analysis \citep[see Table~3
of][]{Abbott:2010}.

\section{Results}
None of the searches yielded evidence of a gravitational wave signal, and upper limits have been placed on signal
strengths. These limits are subject to the uncertainties in the amplitude calibration, as discussed in
Section~\ref{sec:runs}. For the joint results, which combine data from multiple detectors, the sensitivity is often
dominated by the most sensitive instrument. Therefore, we expect the amplitude uncertainty due to calibration
uncertainties to also be dominated by the most sensitive instrument. So, following the calibration error given in
Section~\ref{sec:runs}, below $\sim$50~Hz we have an amplitude uncertainty of $\sim$6\%, and above that we have
uncertainty of $\sim$20\%. The phase uncertainties are small enough to have a negligible contribution to the possible
amplitude uncertainty.

\subsection{Data selection}
\begin{figure*}[!htbp]
\includegraphics[width=1.0\textwidth]{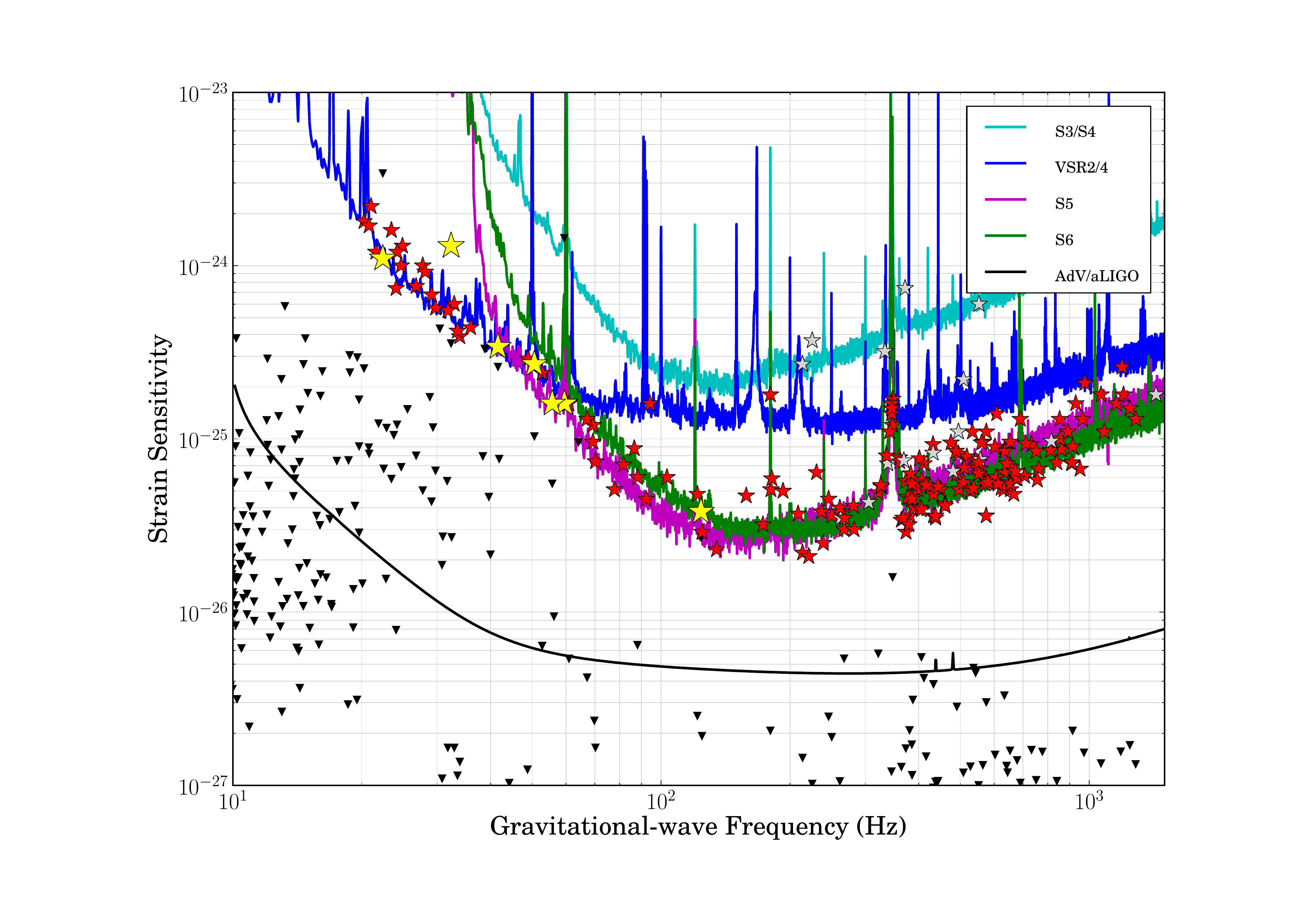}
\caption{The $h_0^{95\%}$ upper limits (given by $\star$) for 195 pulsars from the LIGO and Virgo S3/S4, S5, S6, VSR2,
and VSR4 runs. The curves give estimated relative strain sensitivities of these runs and potential future science runs.
The sensitivities are based on the harmonic mean of the observation time ($T$) weighted one-sided power spectral
densities $S_n$ from all detectors operating during the given run, and are given by $10.8\sqrt{S_n/T}$, where the scale
factor of 10.8 is given in \citet{Dupuis:2005}. The AdV/aLIGO curve assumes a joint analysis of two equally sensitive
advanced LIGO detectors and the advanced Virgo detector operating at their full design sensitivities with one year of
coherent integration \citep[the sensitivity curves are those given in][]{Aasi:2013c}. The $\blacktriangledown$ give the
spin-down limits for all (non-Globular Cluster) pulsars, based on values taken from the ATNF catalog and assuming the
canonical moment of inertia. The $\star$ show the observational upper limits from Tables~\ref{tab:highlights},
\ref{tab:virgoonly} and \ref{tab:s6vsr24table}, with the seven high interest pulsars represented by the larger, lighter
colored stars. Results for pulsars using the previous S3/S4 and S5 data are given by the small lighter colored stars. 
\label{fig:senscurves}}
\end{figure*}

As discussed in Section~\ref{sec:em}, for a few pulsars the electromagnetic observations did not always span the
S6/VSR2,4 runs completely, and some pulsars glitched during the runs. As a result, we deal with these instances
separately. In most cases, we can use all the data coherently, but, in other cases, sections of data must be combined
incoherently. The relative sensitivities of the detectors at the pulsar frequencies also dictate whether we have used
Virgo-only, LIGO-only or Virgo and LIGO data (see Fig.~\ref{fig:senscurves}). For J0537\textminus6910, only the LIGO
data has been used because of its better sensitivity at the corresponding frequency, and results from each inter-glitch
period have been combined incoherently. For J0835\textminus4510 (the Vela pulsar), a glitch occurred just prior to VSR4
and we had no phase-connected timing solution between VSR2 \citep{Abadie:2011} and VSR4 epochs. The VSR2 results and
VSR4 data have therefore been incoherently combined. For J1813\textminus1246, the results from the pre- and post-glitch
periods using all data from S6 and VSR2,4 have been combined incoherently. For J1833\textminus1034, only VSR2 data up to
the time of the observed glitch has been used.

\begin{figure}[!htbp]
\includegraphics[width=0.49\textwidth]{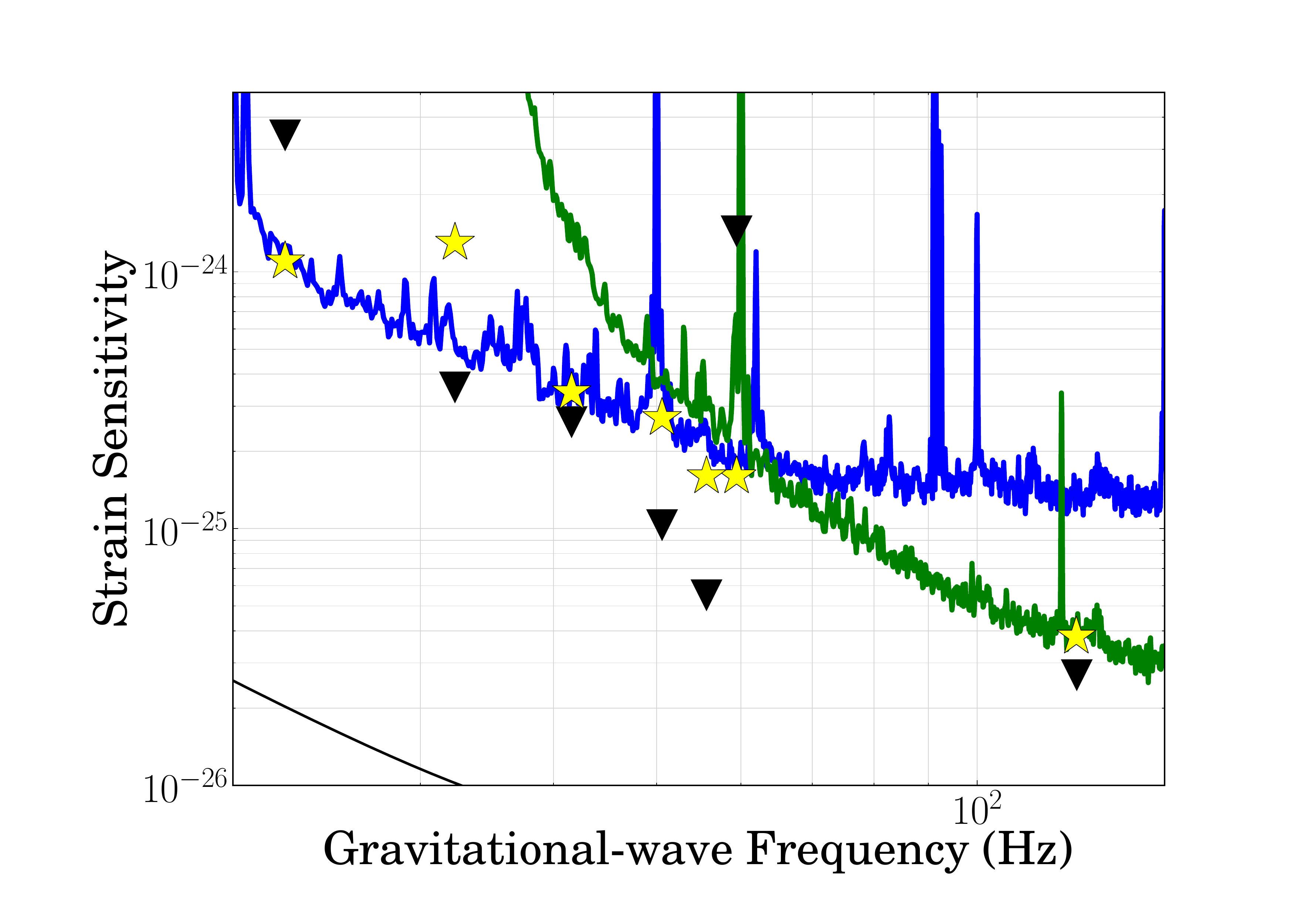}
\caption{A zoomed version of Figure~\ref{fig:senscurves} focusing on the seven high interest pulsars. The outlier at
$\sim$32~Hz is J1833\textminus1034 for which only VSR2 data was used. 
\label{fig:senscurves_zoom}}
\end{figure}

The parameters and results (from the three different analyses discussed in Section~\ref{sec:methods}) for the seven
pulsars highlighted in Section~\ref{sec:em} are given in Tables~\ref{tab:params} and \ref{tab:highlights}, respectively.
Table~\ref{tab:highlights} gives the 95\% upper limit on the \gw amplitude, $h_0^{95\%}$, the equivalent limits (via
equation~\ref{eq:q22}) on the stars fiducial ellipticity, $\varepsilon$, and mass quadrupole moment, $Q_{22}$, the ratio
of the limit to the spin-down limit, $h_0^{95\%}/h_0^{\rm sd}$, and the limit on the \gw luminosity compared to the
total spin-down luminosity. This final value is given in the form of the percentage of the spin-down luminosity required
to produce a \gw at the amplitude limit (it can be seen from equation~\ref{eq:esd} that this is just the square of the
ratio $h_0^{95\%}/h_0^{\rm sd}$). For those pulsars with constrained orientations (see Table~\ref{tab:constraints}) the
results with and without the constraints are also given in Tables~\ref{tab:highlights}. Despite the very tight
constraints given in Table~\ref{tab:constraints} it should be noted that these results would only show minor
differences if the angle ($\iota$ and $\psi$) errors were expanded to several times the given values. This is because
the recovered posterior probability distributions on these parameters are slowly and smoothly varying over their
parameter ranges. A brief discussion of the differences between the upper limits from the different methods is given in
Section~\ref{sec:methods}.

\begin{deluxetable*}{lcccccccc}
\tablewidth{0pt} \tabletypesize{\footnotesize} \tablecaption{The properties of the pulsars of high interest.
\label{tab:params}} \tablehead{\colhead{Pulsar} & \colhead{$\alpha$} & \colhead{$\delta$} & \colhead{\frot (Hz)} &
\colhead{\fgw (Hz)} &
\colhead{$\dot{f}_{\rm rot}$ (Hz/s)} & \colhead{$d$ (kpc)} & \colhead{$\dot{E}$\tablenotemark{$\dagger$} (W)}
& \colhead{$h^{\rm sd}$\tablenotemark{$\dagger$}}} \startdata
J0534+2200 (Crab) & \hms{05}{34}{31}{97} & \dms{22}{00}{52}{07} & 29.72 & 59.44 & $-3.7\ee{-10}$ &
2.0\tablenotemark{a} & $4.6\ee{31}$ & $1.4\ee{-24}$ \\
J0537\textminus6910 (N157B) & \hms{05}{37}{47}{36} & \dms{-69}{10}{20}{40} & 61.97 & 123.94 & $-2.0\ee{-10}$ &
50.0\tablenotemark{b} & $4.9\ee{31}$ & $3.0\ee{-26}$ \\
J0835\textminus4510 (Vela) & \hms{08}{35}{20}{61} & \dms{-45}{10}{34}{88} & 11.19 & 22.39 & $-1.6\ee{-11}$ &
0.29\tablenotemark{c} & $6.9\ee{29}$ & $3.3\ee{-24}$ \\
J1813\textminus1246 & \hms{18}{13}{23}{74} & \dms{-12}{46}{00}{86} & 20.80 & 41.60 & $-7.6\ee{-12}$ &
1.9\tablenotemark{d} & $6.2\ee{29}$ & $2.6\ee{-25}$ \\
J1833\textminus1034 (G21.5\textminus0.9) & \hms{18}{33}{33}{61} & \dms{-10}{34}{16}{61} & 16.16 & 32.33 & $-5.3\ee{-11}$
& 4.8\tablenotemark{e} & $3.4\ee{30}$ & $3.0\ee{-25}$ \\
J1913+1011 & \hms{19}{13}{20}{34} & \dms{10}{11}{23}{11} & 27.85 & 55.70 & $-2.6\ee{-12}$ & 4.5\tablenotemark{f} &
$2.8\ee{29}$ & $2.3\ee{-25}$ \\
J1952+3252 (CTB 80) & \hms{19}{52}{58}{11} & \dms{32}{52}{41}{24} & 25.30 & 50.59 & $-3.7\ee{-12}$ &
3.0\tablenotemark{f} & $3.7\ee{29}$ & $1.0\ee{-25}$
\enddata
\tablenotetext{$\dagger$}{The spin-down luminosity, $\dot{E}$, and spin-down \gw amplitude limit, $h^{\rm
sd}$, both assume a canonical moment of inertia of $I_{zz} = 10^{38}\,{\rm kg}\,{\rm m}^2$.}
\tablenotetext{a}{See Appendix of \citet{Kaplan:2008}.}
\tablenotetext{b}{\citet{Pietrzynski:2013}.}
\tablenotetext{c}{\citet{Dodson:2003}.}
\tablenotetext{d}{This distance is the average of the two estimates from \citet{Wang:2011}, which allow a distance
between $\sim$0.9--3.5~kpc.}
\tablenotetext{e}{\citet{Tian:2008}.}
\tablenotetext{f}{The distance is taken from the ATNF pulsar catalog \citep{Manchester:2005}.}
\end{deluxetable*}

\begin{deluxetable*}{lccccc}
\tabletypesize{\footnotesize} \tablecaption{Upper limits for the high interest pulsars. Limits with constrained
orientations are given in parentheses. \label{tab:highlights}} \tablehead{
\colhead{Analysis} & \colhead{$h_0^{95\%}$} & \colhead{$\varepsilon$} & $Q_{22}$\,(kg\,m$^2$) &
\colhead{$h_0^{95\%}/h_0^{\rm sd}$} & \colhead{$\dot{E}_{\rm gw}/\dot{E}$\,\%}}
\startdata
\cutinhead{J0534+2200 (Crab)}
Bayesian & $1.6\,(1.4)\ee{-25}$ & $8.6\,(7.5)\ee{-5}$ & $6.6\,(5.8)\ee{33}$ & 0.11 (0.10) & 1.2 (1.0) \\
\FG-statistic & $2.3\,(1.8)\ee{-25}$ & $12.3\,(9.6)\ee{-5}$ & $11.6\,(7.4)\ee{33}$ & 0.16 (0.13) & 2.6 (1.7) \\
$5n$-vector & $1.8\,(1.6)\ee{-25}$ & $9.7\,(8.6)\ee{-5}$ & $7.4\,(6.6)\ee{33}$ & 0.12 (0.11) & 1.4 (1.2) \\
\cutinhead{J0537\textminus6910}
Bayesian & $3.8\,(4.4)\ee{-26}$ & $1.2\,(1.4)\ee{-4}$ & $0.9\,(1.0)\ee{34}$ & 1.4 (1.7) & 200 (290) \\
\FG-statistic & $1.1\,(1.0)\ee{-25}$ & $3.4\,(3.1)\ee{-4}$ & $2.6\,(2.4)\ee{34}$ & 4.1 (3.9) & 1700 (1500) \\
$5n$-vector & $4.5\,(6.7)\ee{-26}$ & $1.4\,(2.1)\ee{-4}$ & $1.1\,(1.6)\ee{34}$ & 1.6 (2.4) & 260 (580) \\
\cutinhead{J0835\textminus4510 (Vela)}
Bayesian & $1.1\,(1.0)\ee{-24}$ & $6.0\,(5.5)\ee{-4}$ & $4.7\,(4.2)\ee{34}$ & 0.33 (0.30) & 11 (9.0) \\
\FG-statistic & $4.2\,(9.0)\ee{-25}$ & $2.3\,(4.9)\ee{-4}$ & $1.8\,(3.8)\ee{34}$ & 0.13 (0.27) & 1.7 (7.3) \\
$5n$-vector & $1.1\,(1.1)\ee{-24}$ & $6.0\,(6.0)\ee{-4}$ & $4.7\,(4.7)\ee{34}$ & 0.33 (0.33) & 11 (11) \\ 
\cutinhead{J1813\textminus1246}
Bayesian & $3.4\ee{-25}$ & $3.5\ee{-4}$ & $2.7\ee{34}$ & 1.3 & 170 \\
\FG-statistic & $7.1\ee{-25}$ & $7.4\ee{-4}$ & $5.7\ee{34}$ & 2.7 & 730 \\
$5n$-vector & $4.8\ee{-25}$ & $4.9\ee{-4}$ & $3.8\ee{34}$ & 1.8 & 320 \\
\cutinhead{J1833\textminus1034}
Bayesian & $1.3\,(1.4)\ee{-24}$ & $5.7\,(6.1)\ee{-3}$ & $4.4\,(4.7)\ee{35}$ & 4.3 (4.6) & 1800 (2100) \\
\FG-statistic & $1.2\,(1.2)\ee{-24}$ & $5.2\,(5.2)\ee{-3}$ & $4.0\,(4.0)\ee{35}$ & 3.9 (3.9) & 1500 (1500) \\
$5n$-vector & $1.4\,(2.0)\ee{-24}$ & $6.1\,(8.7)\ee{-3}$ & $4.7\,(6.7)\ee{35}$ & 4.6 (6.6) & 2100 (4400) \\
\cutinhead{J1913+1011}
Bayesian & $1.6\ee{-25}$ & $2.2\ee{-4}$ & $1.7\ee{34}$ & 2.9 & 840 \\
\FG-statistic & $2.9\ee{-25}$ & $4.1\ee{-4}$ & $3.1\ee{34}$ & 5.3 & 2800 \\
$5n$-vector & $2.5\ee{-25}$ & $3.4\ee{-4}$ & $2.7\ee{34}$ & 4.5 & 2000 \\
\cutinhead{J1952+3252}
Bayesian & $2.7\,(2.5)\ee{-25}$ & $3.0\,(2.8)\ee{-4}$ & $2.3\,(2.1)\ee{34}$ & 2.6 (2.5) & 680 (630) \\
\FG-statistic & $6.0\ee{-25}$ & $6.7\ee{-4}$ & $5.1\ee{34}$ & 5.8 & 3400 \\
$5n$-vector & $3.1\,(3.2)\ee{-25}$ & $3.4\,(3.5)\ee{-4}$ & $2.6\,(2.7)\ee{34}$ & 3.0 (3.1) & 900 (960)
\enddata
\tablecomments{Detector calibration errors mean that for pulsars with $f_{\rm gw}$ below and above 50\,Hz (see
Table~\ref{tab:params}) there are $\sim$$6\%$ and $\sim$$20\%$ uncertainties respectively on these limits.}
\end{deluxetable*}

One of the new targets, J1824\textminus2452I \citep[which is an interesting pulsar that is seen to switch between being
accretion and rotation powered][]{Papitto:2013}, had a coherent timing solution that covered 2006, so S5 data from the
LIGO detectors has been reanalyzed for this result. For all other pulsars, we have used only the VSR2 and VSR4 data if
$\fgw < 40$\,Hz, and have coherently combined VSR2, VSR4 and S6 data from H1 and L1 for pulsars with $\fgw > 40$\,Hz.
All the available science mode data (i.e., when the detectors were operating in a stable state) has been used, with
details given in Table~\ref{tab:runs}.

For the 19 pulsars with $\fgw < 40$\,Hz the results can be found in Table~\ref{tab:virgoonly}. Because of their low
frequencies, none of these pulsars had been targeted before.

Results for pulsars with $\fgw > 40$\,Hz using S6 and VSR2,4 are shown in Table~\ref{tab:s6vsr24table}. Distances
to pulsars in Terzan 5 (with designations J1748\textminus2446) are assumed to be 5.5\,kpc \citep{Ortolani:2007} rather
than the value of 8.7\,kpc given in the ATNF catalog, and distances to the pulsars in M28 (with designations
J1824\textminus2452) are assumed to be 5.5\,kpc \citep{Harris:1996} rather than the distance of 4.9\,kpc given in
\citet{Abbott:2010}. Unless otherwise specified in the table for all other pulsars we use the distance values given by
the {\tt DIST} value in the ATNF catalog \citep{Manchester:2005}, which generally are dispersion measure calculations
from the electron density distribution model of \citet{Taylor:1993}. For the 16 pulsars where new timing solutions were
not available during the most recent runs (see Section~\ref{sec:em}), we include the results from the LIGO S3/S4
\citep{Abbott:2007a} and S5 analysis \citep{Abbott:2010}.

\begin{figure*}[!htbp]
\begin{tabular}{cc}
\includegraphics[width=0.45\textwidth]{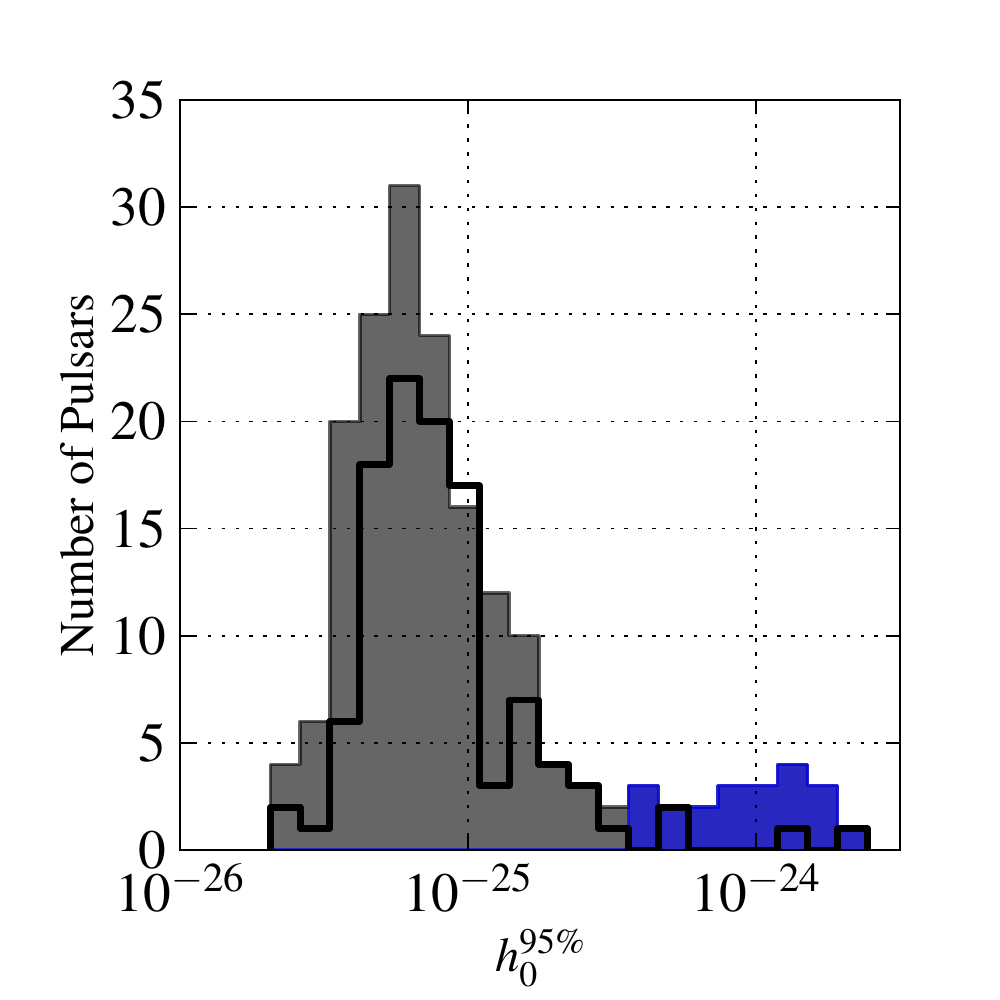} & \includegraphics[width=0.45\textwidth]{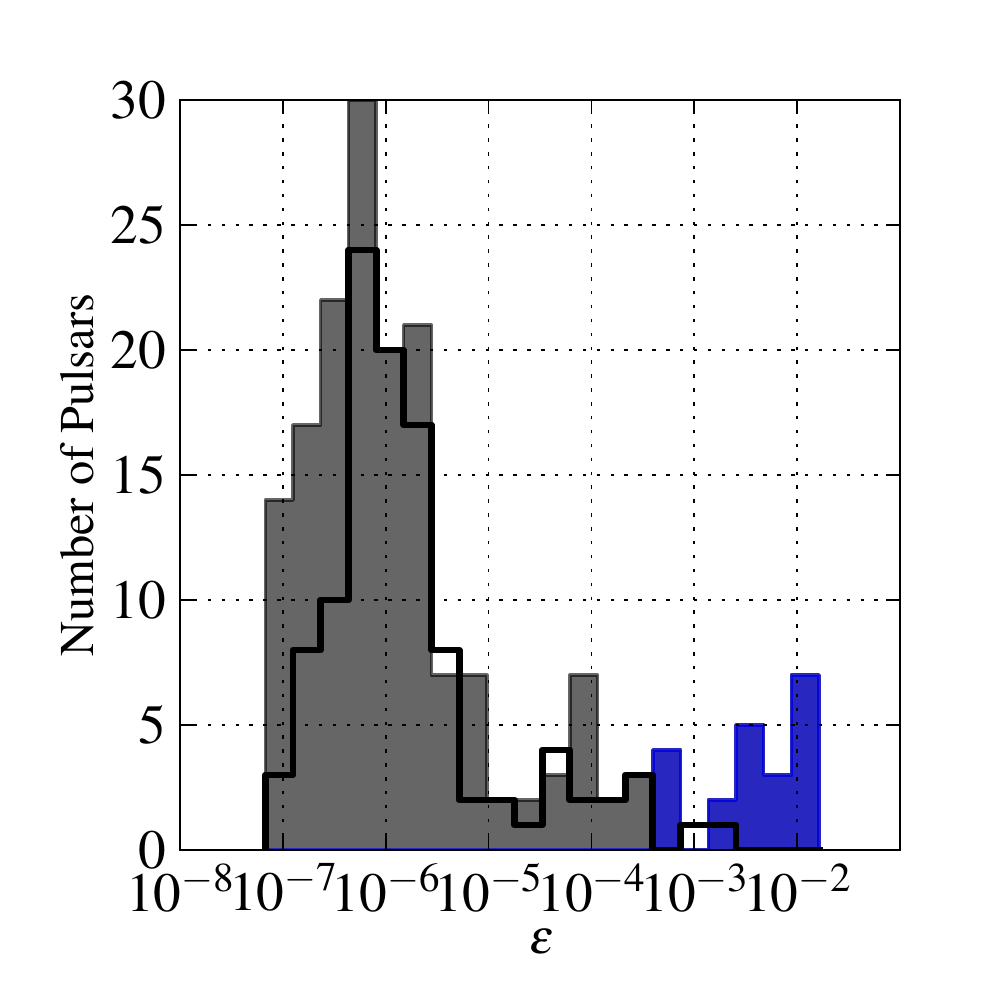} \\
\includegraphics[width=0.45\textwidth]{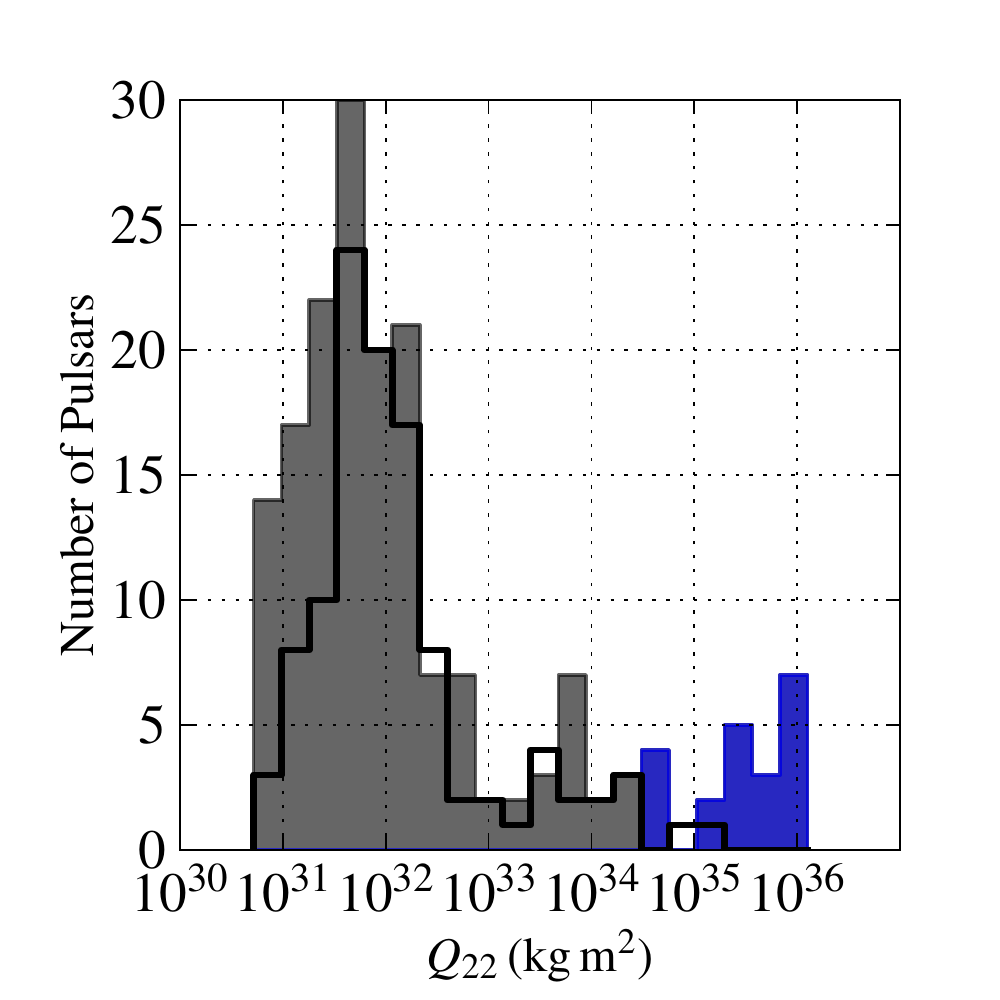} & \includegraphics[width=0.45\textwidth]{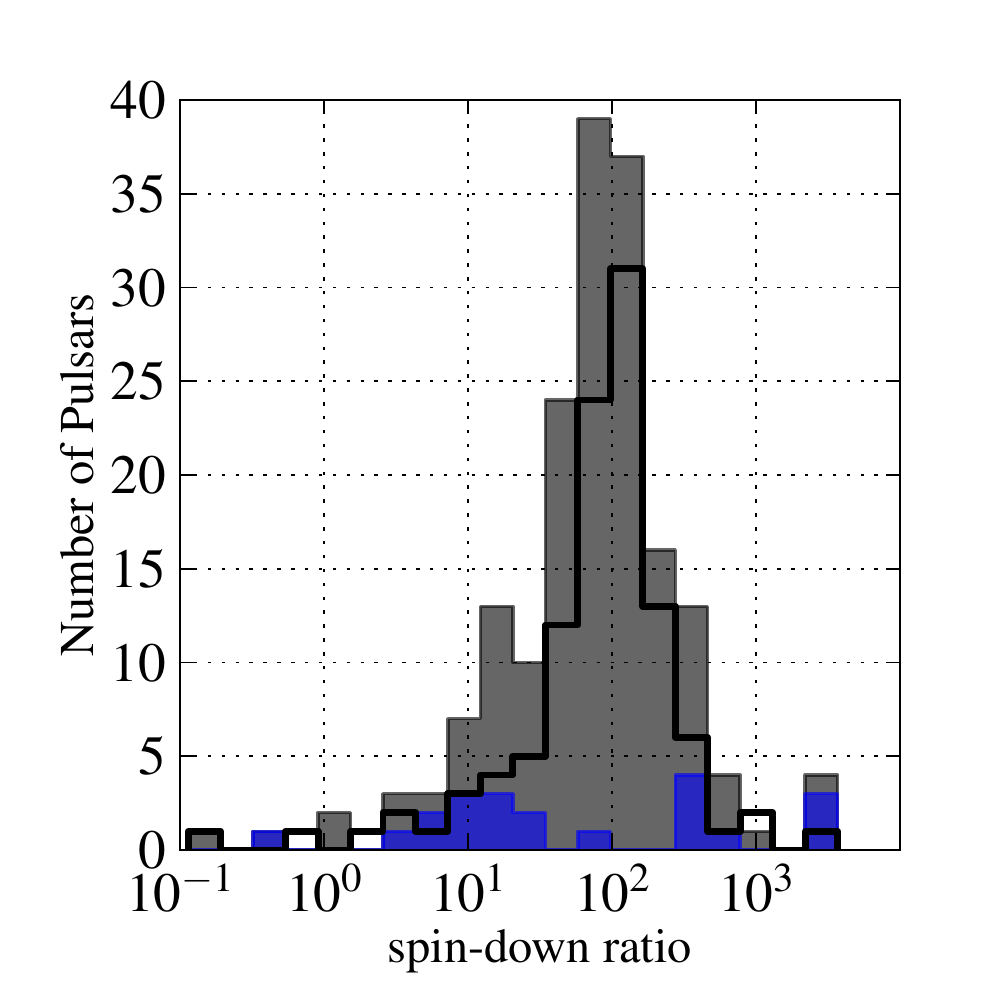}
\end{tabular}
\caption{The upper limits in histogram form for all pulsars for $h_0$, $\varepsilon$, $Q_{22}$ and the spin-down limit
ratio. The grey shaded area represents results from the S6/VSR2,4 analysis combining all detectors, the blue shaded area
represents results from the VSR2,4-only analyses. These also contain the seven high interest pulsars for which the
Bayesian method values have been plotted based on no assumptions about the pulsar orientations. Previous upper limits
from the S5 analysis are given by the unfilled histogram. \label{fig:reshists}}
\end{figure*}

The \gw amplitude upper limits as a function of frequency are plotted in Fig.~\ref{fig:senscurves} and
Fig.~\ref{fig:senscurves_zoom} (showing a version just containing the seven high interest pulsars), which also show
bands giving the expected sensitivity of the analysis. The upper limits in histogram form for all pulsars can be seen in
Fig.~\ref{fig:reshists}. The histograms show that the distribution of $h_0$ upper limits is peaked just below
$10^{-25}$, corresponding to equivalent peaks on $\varepsilon$ and $Q_{22}$ of $\sim$$10^{-6}$ and
$\sim$$10^{-32}$\,kg\,m$^2$. The spin-down limit ratios shows that we are within a factor of 100 for just over half of
the pulsars. It is interesting to see that due to the shape of the detector sensitivity curves the lower frequency young
pulsars (analysed only with Virgo data) have the highest amplitude limits, but as several have high spin-down
luminosities they have an approximately uniform spread in spin-down limit ratios.

\section{Discussion}\label{sec:discussion}
We have seen no credible evidence for \gw emission from any known pulsar, but have been able to place upper limits on
the \gw amplitude from an unprecedented number of pulsars. In this work we have produced entirely new results for 73
pulsars and updated the results of previously searches for 106 pulsars, with results from a further 16 from previous
analyses being reproduced here. A total of 195 pulsars have now been targeted over the lifetimes of the first generation
of interferometric \gw detectors.

\subsection{Quadrupole estimates}
As discussed in the introduction, we have targeted the gravitational-wave signature of the time-varying $l=m=2$
quadrupole moment. There is great uncertainty, however, as to whether neutron stars can form and sustain sufficient
elastic deformations to give an observable quadrupole, and this, in turn, makes it difficult to model a realistic source
population. The recent work by \citet{Johnson-McDaniel:2012} \citep[also see e.g.][]{Owen:2005, Pitkin:2011} on the
maximum sustainable quadrupole for a variety of neutron star equations of state indicates that relatively large
quadrupoles can indeed be sustained. \citet{Johnson-McDaniel:2012} find that solid quark stars could sustain quadrupoles
of up to $10^{37}$\,kg\,m$^2$ (or fiducial ellipticities of order $0.1$), hybrid stars could sustain quadrupoles of up
to $10^{35}$\,kg\,m$^2$ (or fiducial ellipticities of order $1\ee{-3}$), while for normal neutron stars the stiffest
equations of state allow quadrupoles of $\sim$$1\ee{33}$\,kg\,m$^2$ (or fiducial ellipticities of $\sim$$1\ee{-5}$). It
is worth noting that these are maximum allowable quadrupoles, and it is still unknown whether they are realized in
nature for reasons described in \citet{Abbott:2007c}.

A mass quadrupole may also be generated by distortional pressure from the star's magnetic field
\citep[see e.g.][]{Bonazzola:1996, Cutler:2002, Ciolfi:2010}. The external dipole field of a pulsar is usually estimated
from its rotational spin-down, assuming this is due to magnetic dipole radiation (equivalent to the \gw spin-down limit
that we define). As discussed in Section~\ref{sec:pulsars}, this gives external surface dipole field strengths of
$\sim$$10^9$\,G for MSPs and $\sim$$10^{12}$\,G for normal pulsars. Internal fields of this magnitude are too small to
induce mass quadrupoles that would be currently observable, but the field strengths of some magnetars are at a suitable
level (though rotating too slowly to be detectable sources for ground-based \gw detectors). Unfortunately, internal
field strengths and configurations are not well understood, and the mechanisms for burying fields beneath the surface
are uncertain. Studies of one young pulsar with a braking index of $n \approx 1$ \citep{Espinoza:2011} may point towards
an evolving and increasing external magnetic field, with an internal field leaking out over time, but recently other
mechanisms have been proposed to explain the evolution of the field that do not rely on an increasing magnetic field
\citep[e.g.][]{Ho:2012, Caliskan:2013}. \citet{Mastrano:2012} discuss the prospects of constraining field strength and
configuration for recycled MSPs using \gw data. This is also discussed in \citet{Pitkin:2011}, who shows limits that
could be obtained on fully poloidal or toroidal field configurations. Further estimates of the quadrupoles that can be
generated by internal magnetic fields for a given equation of state are given in \citet{Haskell:2007bh, Haskell:2009,
Akgun:2007ph}.

\subsection{High interest pulsars}
For the seven high interest pulsars the results are all close to (or beat) the spin-down limits. In particular, our
upper limits are significantly below the spin-down limit for the Crab and Vela pulsars, further improving over past
results. The mass quadrupole limits are generally within $10^{34}$--$10^{35}$\,kg\,m$^2$, with the Crab pulsar upper
limit slightly lower at $\sim$$7\ee{33}$\,kg\,m$^2$. Therefore, for these stars to emit \gws at current sensitivities
the emission would most likely have to come from a quark star or one with a hybrid core, whilst the Crab pulsar is about
an order of magnitude above the maximum quadrupoles expected for purely crustal emission. However, for advanced
detectors the sensitivity for Crab pulsar would be consistent with most optimistic predictions for {\it normal} neutron
stars. For J0537\textminus6910, which has a quadrupole limit close to the Crab pulsar, future prospects may not be so
good for reaching the most optimistic prediction for {\it normal} neutron stars. This is due to the requirement for
phase coherent timing, which for these analyses relied on the no-longer-operational RXTE.

For the Crab and Vela pulsars, our results now limit the \gw emission to contribute $\lesssim 1\%$ and $\lesssim 10\%$
of their respective spin-down luminosities, with an improvement of about a factor of 4 for Vela with respect to previous
results. These limits are compatible with the observed braking indices of the pulsars, which are $n=2.51$ and
$n\approx1.4$ respectively \citep[see e.g.][]{Palomba:2000}.

Given various assumptions about the magnetic field discussed above, our results constrain the internal field of the Crab
pulsar to be less than $\sim$$10^{16}$\,G \citep[e.g.][]{Cutler:2002}. For the other high interest pulsars, the limits
on the magnetic field would be even higher than this, so we have not included them here.

\citet{Johnson-McDaniel:2013} relates the limits on the $l=m=2$ quadrupole moment from the \gw emission to the physical
surface deformation of a star for a variety of equations of state, which can be compared to the oblateness due to
rotation (note that there is no particular reason to expect a relation between these quantities). His results showed
that previous \gw limits for the Crab pulsar constrained the surface deformation from the $l=m=2$ quadrupole to be well
below the rotational deformation for all equations of state and neutron star masses. Our new results slightly improve
these limits, with the physical surface deformation limited to less than $\sim$$30$\,cm, maximized over masses and
equations of state. For the Vela pulsar, our new results limit non-axisymmetric quadrupole deformations to be $\lesssim
100$\,cm, which is smaller than the expected rotational oblateness for equations of state with large radii.

For PSR J0537\textminus6910, the quality of S6 data at the corresponding frequency was relatively poor, and the upper
limits are no better than those produced during S5 \citep{Abbott:2010}. If this pulsar were, however, at the upper end
of the moment of inertia range ($\sim$$3\ee{38}$\,kg\,m$^2$) the spin-down limit would be increased by a factor of
$\sim$1.7, and we would now fall below it\footnote{The distance to the Large Magellanic Cloud is known to $\sim$2\%
\citep{Pietrzynski:2013}, so does not significantly contribute to the uncertainty on the spin-down limit.}.

\subsection{Other highlights}
Several other pulsar upper limits are within a factor of 10 of their spin-down limits. For the MSPs, three upper limits
are within a factor of ten of the spin-down limit: J1045\textminus4509, a factor of 6; J1643\textminus1224, a factor of
10; and J2124\textminus3358, also a factor of 10. The upper limit that corresponds to the smallest ellipticity/mass
quadrupole is from J2124\textminus3358 with $\varepsilon = 6.7\ee{-8}$ and $Q_{22} = 5.2\ee{30}$\,kg\,m$^2$. Although
this value is currently above the spin-down limit, it is well within allowable maximum deformations for all neutron star
equations of state \citep[see e.g.][]{Pitkin:2011}. The \gw spin-down limits for these pulsars require quadrupoles that are
well within reasonable theoretical ranges, so they will make intriguing targets for the advanced generation of
detectors.

For the young pulsars only targeted with Virgo VSR2 and VSR4 data a further five are within a factor of ten of the
spin-down limit (see Table~\ref{tab:virgoonly}). All of these would be required to have an exotic equation of state to
be observed at around their spin-down limits in future detectors. 

\subsection{Future prospects}
The search results described in this paper assume that the pulsar gravitational-wave phase evolution is very well known
and tied very closely to the observed electromagnetic phase. However, precession \citep[e.g.][]{Zimmerman:1979,
Jones:2002} or other models \citep{Jones:2010} could give emission at both the rotation frequency and twice the rotation
frequency. Additionally, as discussed in \citet{Abbott:2008}, emission may be offset from the electromagnetic phase
model. We therefore will be applying methods to search for \gws from known pulsars at multiple harmonics and with narrow
bandwidths around the observed electromagnetic values in archival and future datasets.

We look forward to the era of the {\it advanced} LIGO (aLIGO) \citep{Harry:2010} and Virgo (AdV)
\citep{Acernese:2009, Accadia:2012} detectors \citep[see][for estimates of the aLIGO and AdV observation schedule and
sensitivity evolution]{Aasi:2013c}, as well as the KAGRA detector \citep{Somiya:2012}. Ongoing radio pulsar surveys are
discovering new objects that will be targeted with future detectors. Currently, the High Time Resolution Universe survey
with the Parkes and Effelsberg telescopes \citep{Keith:2010} has discovered 29 new MSPs \citep{Keith:2013, Ng:2013} and
could discover up to $\sim$75 once complete. The high sensitivity Arecibo PALFA survey is discovering new pulsars
\citep{Lazarus:2013} and making use of distributed computing through Einstein@home \citep{Allen:2013}. The Green Bank
Drift-scan survey and the Green Bank North Celestial Cap survey are also discovering new and interesting sources
\citep{Lynch:2013}. Many interesting high energy pulsars, undetectable in the radio frequency band, are also being
detected by the {\it Fermi} Large Area Telescope \citep{SazParkinson:2013}. {\it Fermi} is also providing targets to
facilitate radio searches which are finding many new MSPs. In addition, new analyses of archive data, such as using
Einstein@home to search through Parkes Multi-beam Pulsar Survey data, are still yielding new results
\citep{Knispel:2013}. In the near future, there are exciting prospects from the Low Frequency Array (LOFAR), which could
detect the majority of radio pulsars within $\sim$2\,kpc, giving of order 1000 new pulsars \citep{vanLeeuwen:2010,
Stappers:2011}, and perform deep searches for pulsars in globular clusters.

Finally, we should emphasize that known pulsar searches are not the only searches looking for \gws from rotating,
galactic neutron stars. There have been, or are under way, several directed searches looking for sources of unknown
frequency and spin-down in particular objects e.g.\ globular clusters, supernova remnants \citep[e.g.][]{Abadie:2010,
Chung:2011}, the Galactic center \citep{Aasi:2013b}, and low-mass X-ray binaries. There are also several semi-coherent,
all-sky, wide-frequency band searches \citep[e.g.][]{Abadie:2012, Aasi:2013}. Very similar pipelines will be used
during the advanced detector era, yielding signal candidates, performing follow-ups and, in case of detection, source
parameter estimation.

\acknowledgements The authors gratefully acknowledge the support of the United States National Science Foundation for
the construction and operation of the LIGO Laboratory, the Science and Technology Facilities Council of the United
Kingdom, the Max-Planck-Society, and the State of Niedersachsen/Germany for support of the construction and operation of
the GEO600 detector, and the Italian Istituto Nazionale di Fisica Nucleare and the French Centre National de la
Recherche Scientifique for the construction and operation of the Virgo detector. The authors also gratefully acknowledge
the support of the research by these agencies and by the Australian Research Council, the International Science Linkages
program of the Commonwealth of Australia, the Council of Scientific and Industrial Research of India, the Istituto
Nazionale di Fisica Nucleare of Italy, the Spanish Ministerio de Econom\'ia y Competitividad, the Conselleria d'Economia
Hisenda i Innovaci\'o of the Govern de les Illes Balears, the Foundation for Fundamental Research on Matter supported by
the Netherlands Organisation for Scientific Research, the Polish Ministry of Science and Higher Education, the FOCUS
Programme of Foundation for Polish Science, the Royal Society, the Scottish Funding Council, the Scottish Universities
Physics Alliance, the National Aeronautics and Space Administration, OTKA of Hungary, the Lyon Institute of Origins
(LIO), the National Research Foundation of Korea, Industry Canada and the Province of Ontario through the Ministry of
Economic Development and Innovation, the National Science and Engineering Research Council Canada, the Carnegie Trust,
the Leverhulme Trust, the David and Lucile Packard Foundation, the Research Corporation, and the Alfred P.\ Sloan
Foundation.
% Acknolegdement for Nancay
The Nan\c{c}ay Radio Observatory is operated by the Paris Observatory, associated with the French Centre National de la
Recherche Scientifique.
LIGO Document No.\ LIGO-P1200104.

\appendix

\section{Ephemeris for J0537\textminus6910}\label{app:j0537}
Over the span of S6, VSR2, and VSR4 RXTE made observations of J0537\textminus6910. It was observed to glitch four times
during this period and phase connected ephemerides were produced for each inter-glitch segment. These ephemerides, given
in Table~\ref{tab:J0537-6910ephem}, use a DE200 sky position of $\alpha =$ \hms{05}{37}{47}{36} and $\delta =$
\dms{-69}{10}{20}{4} \citep{Wang:2001}.
\begin{deluxetable*}{ccccc}
\tablewidth{0pt} \tabletypesize{\footnotesize} \tablecaption{RXTE ephemerides for J0537\textminus6910 during the
period of S6, VSR2, and VSR4. \label{tab:J0537-6910ephem}} \tablehead{\colhead{EM Observation Span (MJD and date)} &
\colhead{Epoch $t_0$ (MJD$_{\rm TDB}$)} & \colhead{\frot (Hz)} & \colhead{$\dot{f}_{\rm rot}$ ($10^{-10}$\,Hz/s)} &
\colhead{$\ddot{f}_{\rm rot}$ ($10^{-20}$\,Hz/s$^2$)}} \startdata
54897 (2009 Mar 7) -- 55041 (2009 Jul 29) & 54966.4266616022 & 61.9765822442 & $-1.9948300$ & $0.960$ \\
55045 (2009 Aug 2) -- 55182 (2009 Dec 17) & 55140.0377727133 & 61.9736036203 & $-1.9939924$ & $2.090$ \\
55185 (2009 Dec 20) -- 55263 (2010 Mar 8) & 55221.0562912318 & 61.9722204755 & $-1.9951973$ & $2.090$ \\
55275 (2010 Mar 20) -- 55445 (2010 Sep 6) & 55313.6488838244 & 61.9706584446 & $-1.9953330$ & $0.673$ \\
55458 (2010 Sep 19) -- 55503 (2010 Nov 3) & 55475.6859208615 & 61.9678758925 & $-1.9952574$ & $0.673$
\enddata
\end{deluxetable*}

\bibliography{S6KnownPulsarPaper}

% $Id: virgotable.tex,v 1.13 2013/09/30 13:42:06 mpitkin Exp $

%\documentclass[preprint,12pt]{aastex}
%\documentclass{emulateapj}

%\usepackage{amsmath}
% The rotate command for deluxetable does not work in emulateapj, so use a 
% landscape mode instead
%\usepackage{lscape}

%\begin{document}

%\clearpage
%\LongTables % table can span multiple pages
%\begin{landscape}
\begin{deluxetable*}{l c c c c c c c c c}
%\rotate
\tabletypesize{\scriptsize}
\tablewidth{0pt}
\tablecaption{Limits on the gravitational wave amplitude for known pulsars with $\fgw < 40$\,Hz using VSR2,4
data\label{tab:virgoonly}}
\tablehead{\colhead{Pulsar} 
& \colhead{$f_{\rm rot}$ (Hz)} 
& \colhead{$f_{\rm gw}$ (Hz)} 
& \colhead{$\dot{f}_{\rm rot}$ (Hz/s)} 
& \colhead{$d$ (kpc)} 
& \colhead{$h_0^{\rm sd}$} 
& \colhead{$h_0^{95\%}$}
& \colhead{$\varepsilon$}
& \colhead{$Q_{22}$ (kg\,m$^2$)}
& \colhead{$h_0^{95\%}/h_0^{\rm sd}$}
}
\startdata
J0106+4855 
 & 12.03
 & 24.05
 & \ensuremath{-6.2\!\times\!10^{-14}}
 & 7.3
 & \ensuremath{7.9\!\times\!10^{-27}}
& \ensuremath{7.4\!\times\!10^{-25}} 
& \ensuremath{8.9\!\times\!10^{-3}} 
& \ensuremath{6.9\!\times\!10^{35}} 
& 94 
\\
J0609+2130 
 & 17.95
 & 35.91
 & \ensuremath{-7.6\!\times\!10^{-17}}
 & 1.8
 & \ensuremath{9.1\!\times\!10^{-28}}
& \ensuremath{4.4\!\times\!10^{-25}} 
& \ensuremath{5.9\!\times\!10^{-4}} 
& \ensuremath{4.6\!\times\!10^{34}} 
& 490 
\\
J1528\textminus3146 
 & 16.44
 & 32.88
 & \ensuremath{-6.7\!\times\!10^{-17}}
 & 1.0
 & \ensuremath{1.6\!\times\!10^{-27}}
& \ensuremath{6.0\!\times\!10^{-25}} 
& \ensuremath{5.2\!\times\!10^{-4}} 
& \ensuremath{4.0\!\times\!10^{34}} 
& 360 
\\
J1718\textminus3825 
 & 13.39
 & 26.78
 & \ensuremath{-2.4\!\times\!10^{-12}}
 & 4.2
 & \ensuremath{8.0\!\times\!10^{-26}}
& \ensuremath{7.6\!\times\!10^{-25}} 
& \ensuremath{4.2\!\times\!10^{-3}} 
& \ensuremath{3.3\!\times\!10^{35}} 
& 9.5 
\\
J1747\textminus2958 
 & 10.12
 & 20.24
 & \ensuremath{-6.3\!\times\!10^{-12}}
 & 2.5
 & \ensuremath{2.6\!\times\!10^{-25}}
& \ensuremath{1.8\!\times\!10^{-24}} 
& \ensuremath{1.0\!\times\!10^{-2}} 
& \ensuremath{7.9\!\times\!10^{35}} 
& 7.0 
\\
J1748\textminus2446J 
 & 12.45
 & 24.89
 & \ensuremath{-2.0\!\times\!10^{-16}}
 & 5.5
 & \ensuremath{5.8\!\times\!10^{-28}}\tablenotemark{$\ddagger$}
& \ensuremath{1.3\!\times\!10^{-24}} 
& \ensuremath{1.1\!\times\!10^{-2}} 
& \ensuremath{8.3\!\times\!10^{35}} 
& 2200 
\\
J1753\textminus1914 
 & 15.88
 & 31.77
 & \ensuremath{-4.9\!\times\!10^{-16}}
 & 2.8
 & \ensuremath{1.6\!\times\!10^{-27}}
& \ensuremath{5.5\!\times\!10^{-25}} 
& \ensuremath{1.4\!\times\!10^{-3}} 
& \ensuremath{1.1\!\times\!10^{35}} 
& 340 
\\
J1753\textminus2240 
 & 10.51
 & 21.02
 & \ensuremath{-6.9\!\times\!10^{-17}}
 & 3.5
 & \ensuremath{6.0\!\times\!10^{-28}}
& \ensuremath{2.2\!\times\!10^{-24}} 
& \ensuremath{1.6\!\times\!10^{-2}} 
& \ensuremath{1.3\!\times\!10^{36}} 
& 3700 
\\
\enddata
\tablenotetext{$\dagger$}{The pulsar's spin-down is corrected for proper motion effects.}
\tablenotetext{$\ddagger$}{The pulsar's spin-down is calculated using a characteristic spin-down age of $10^9$ years.}
\tablecomments{Detector calibration errors mean that for these pulsars there are $\sim$$6\%$ uncertainties on these
limits.}
\end{deluxetable*}
%\clearpage
%\end{landscape}
%\end{document}

% $Id: s6vsr24table.tex,v 1.20 2013/09/30 13:42:05 mpitkin Exp $

%\documentclass[preprint,12pt]{aastex}
%\documentclass{emulateapj}

%\usepackage{amsmath}
% The rotate command for deluxetable does not work in emulateapj, so use a
% landscape mode instead
%\usepackage{lscape}

%\begin{document}

\clearpage

% shift the horizontal offset, so that the table is centred (unfortunately as a by product this shifts the page
% numbering!)
\advance\hoffset by -13mm

\LongTables % table can span multiple pages
%\begin{landscape}
\begin{deluxetable*}{l c c c c c c c c c c}
%\rotate
\tabletypesize{\footnotesize}
%\tablewidth{0pt}
\tablecaption{Limits on the gravitational wave amplitude for known pulsars with $\fgw >40$\,Hz\label{tab:s6vsr24table}}
\tablehead{\colhead{Pulsar} 
& \colhead{$f_{\rm rot}$ (Hz)} 
& \colhead{$f_{\rm gw}$ (Hz)} 
& \colhead{$\dot{f}_{\rm rot}$ (Hz/s)} 
& \colhead{$d$ (kpc)} 
& \colhead{$h_0^{\rm sd}$} 
& \colhead{prior $h_0^{95\%}$} 
& \colhead{S6/VSR2,4 $h_0^{95\%}$}
& \colhead{$\varepsilon$}
& \colhead{$Q_{22}$ (kg\,m$^2$)}
& \colhead{$h_0^{95\%}/h_0^{\rm sd}$}
}
\startdata
J0023+0923
 & 327.85
 & 655.69
 & \ensuremath{-1.3\!\times\!10^{-15}}
 & 0.9
 & \ensuremath{1.7\!\times\!10^{-27}}
 &\nodata
& \ensuremath{6.9\!\times\!10^{-26}} 
& \ensuremath{1.5\!\times\!10^{-7}} 
& \ensuremath{1.1\!\times\!10^{31}} 
& 41 
\\
J0024\textminus7204C 
 & 173.71
 & 347.42
 & \ensuremath{1.6\!\times\!10^{-15}}
 & 4.0
 & \ensuremath{6.1\!\times\!10^{-28}}\tablenotemark{$\dagger$}
 & \ensuremath{5.8\!\times\!10^{-25}}
& \ensuremath{1.7\!\times\!10^{-25}} 
& \ensuremath{5.5\!\times\!10^{-6}} 
& \ensuremath{4.2\!\times\!10^{32}} 
& 290 
\\
J0024\textminus7204D 
 & 186.65
 & 373.30
 & \ensuremath{1.6\!\times\!10^{-16}}
 & 4.0
 & \ensuremath{1.8\!\times\!10^{-28}}\tablenotemark{$\dagger$}
 & \ensuremath{4.5\!\times\!10^{-26}}
& \ensuremath{2.9\!\times\!10^{-26}} 
& \ensuremath{7.8\!\times\!10^{-7}} 
& \ensuremath{6.1\!\times\!10^{31}} 
& 160 
\\
J0024\textminus7204E 
 & 282.78
 & 565.56
 & \ensuremath{-7.8\!\times\!10^{-15}}
 & 4.0
 & \ensuremath{1.1\!\times\!10^{-27}}\tablenotemark{$\dagger$}
 & \ensuremath{9.9\!\times\!10^{-26}}
& \ensuremath{9.7\!\times\!10^{-26}} 
& \ensuremath{1.1\!\times\!10^{-6}} 
& \ensuremath{8.9\!\times\!10^{31}} 
& 92 
\\
J0024\textminus7204F 
 & 381.16
 & 762.32
 & \ensuremath{-9.3\!\times\!10^{-15}}
 & 4.0
 & \ensuremath{9.9\!\times\!10^{-28}}\tablenotemark{$\dagger$}
 & \ensuremath{8.8\!\times\!10^{-26}}
& \ensuremath{6.7\!\times\!10^{-26}} 
& \ensuremath{4.4\!\times\!10^{-7}} 
& \ensuremath{3.4\!\times\!10^{31}} 
& 67 
\\
J0024\textminus7204G 
 & 247.50
 & 495.00
 & \ensuremath{2.6\!\times\!10^{-15}}
 & 4.0
 & \ensuremath{6.6\!\times\!10^{-28}}\tablenotemark{$\dagger$}
 & \ensuremath{9.9\!\times\!10^{-26}}
& \ensuremath{8.2\!\times\!10^{-26}} 
& \ensuremath{1.3\!\times\!10^{-6}} 
& \ensuremath{9.8\!\times\!10^{31}} 
& 120 
\\
\enddata
\tablenotetext{$\dagger$}{The pulsar's spin-down is corrected for proper motion effects.}
\tablenotetext{$\ddagger$}{The pulsar's spin-down is calculated using a characteristic spin-down age of $10^9$ years.}
\tablenotetext{\#}{The pulsar's distance is calculated using the NE2001 model of \citet{Cordes:2002}.}
\tablenotetext{a}{Results from S3/S4 \citep{Abbott:2007a}}
\tablenotetext{b}{Results from S5 \citep{Abbott:2010}.}
\tablenotetext{c}{New result from S5 using data covering 2006.}
\tablenotetext{d}{Recently discovered pulsar with timing solution from after the end of S6 or VSR4. As this is an MSP
the timing solution should extrapolate back well over the search period and it is unlikely that any glitches occurred.}

\tablecomments{Detector calibration errors mean that for pulsars with $f_{\rm gw}$ below and above 50\,Hz there are
$\sim$$6\%$ and $\sim$$20\%$ uncertainties on these limits respectively.}

\end{deluxetable*}

\clearpage

% reset the horizontal offset
\advance\hoffset by 13mm

%\clearpage
%\end{landscape}
%\end{document}

\end{document}